\documentclass{llncs}

\usepackage{natbib}
\usepackage{graphicx}

\usepackage{fancyhdr}
\usepackage{lastpage}
 
\begin{document}

\title{A short review and primer on electrodermal activity in human computer interaction applications}
\author{Benjamin Cowley\inst{1,2} \and Jari Torniainen\inst{1}}
\institute{Quantitative Employee unit, Finnish Institute of Occupational Health,\\
\email{benjamin.cowley@ttl.fi},\\
POBox 40, Helsinki, 00250, Finland
\and
Cognitive Brain Research Unit, Institute of Behavioural Sciences, University of Helsinki, Helsinki, Finland}

\maketitle              % typeset the title of the contribution

\begin{abstract}
The application of psychophysiology in human-computer interaction is a growing field with significant potential for future smart personalised systems. Working in this emerging field requires comprehension of an array of physiological signals and analysis techniques. 

One of the most widely used signals is electrodermal activity, or EDA, also known as galvanic skin response or GSR. This signal is commonly used as a proxy for physiological arousal, but recent advances of interpretation and analysis suggest that traditional approaches should be revised. We present a short review on the application of EDA in human-computer interaction. 

This paper aims to serve as a primer for the novice, enabling rapid familiarisation with the latest core concepts. We put special emphasis on everyday human-computer interface applications to distinguish from the more common clinical or sports uses of psychophysiology.

This paper is an extract from a comprehensive review of the entire field of ambulatory psychophysiology, including 12 similar chapters, plus application guidelines and systematic review. Thus any citation should be made using the following reference:

{\parshape 1 2cm \dimexpr\linewidth-1cm\relax
B. Cowley, M. Filetti, K. Lukander, J. Torniainen, A. Henelius, L. Ahonen, O. Barral, I. Kosunen, T. Valtonen, M. Huotilainen, N. Ravaja, G. Jacucci. \textit{The Psychophysiology Primer: a guide to methods and a broad review with a focus on human-computer interaction}. Foundations and Trends in Human-Computer Interaction, vol. 9, no. 3-4, pp. 150--307, 2016.
\par}


\end{abstract}

\section{Introduction}
`Electrodermal activity' (EDA) is a general term used to describe changes in the electrical properties of the skin resulting from autonomic nervous system functions \citep{Dawson2000}. These fluctuations are caused by activation of sweat glands that are controlled by the sympathetic nervous system, which autonomously regulates the mobilisation of the human body for action. Furthermore, skin conductivity is not influenced by parasympathetic activation. Therefore, EDA can be considered to act as an indicator of both psychological and physiological arousal and, by extension, as a measure of cognitive and emotional activity \citep{Dawson2000,boucs12}.

%Electrodermal activity (EDA) is a general term used to describe changes in the electrical properties of the skin resulting from autonomic nervous system functions \citep{Dawson2000}. These fluctuations are caused by activating sweat glands which are controlled by the sympathetic nervous system that autonomously regulates the mobilisation of the human body for action. Furthermore, skin conductivity is not influenced by parasympathetic activation. Thus, EDA can be considered to act as an indicator of both psychological and physiological arousal and, by extension, as a measure of cognitive and emotional activity \citep{Dawson2000,boucs12}.

\section{Background}

EDA has been investigated for well over 100 years, with a number of changes having occurred in the method and the understanding of the phenomenon. Terms have changed accordingly, though `galvanic skin response' is still commonly in use, which can be confusing; instead, one should use the modern terminology, as outlined in \citet{Boucsein2012}: 
\begin{quote}
``[The] first two letters refer to the method of measurement \ldots SP for skin potential, SR for skin resistance, SC for skin conductance, SZ for skin impedance, and SY for skin admittance. The third letter refers to level (L) or response (R)''.
\end{quote}
These terms are derived from the methods employed to detect changes in the electrical properties of the skin, which are the following: the passive measurement of electrical potential difference, or the \textit{endosomatic} method, and active \textit{exosomatic} measurement, wherein either alternating current (AC) or direct current (DC) is passed between two electrodes to measure the skin's \textit{conductivity}, the reciprocal of its \textit{resistance}. In this section, we refer to the latter method, as it is the more widely used (to our knowledge). For full details on these methods, especially how to obtain the slightly more complicated SZ and SY terms, see the work of Boucsein and colleagues.

In the literature, EDA has most often been taken as a measure of arousal \citep{Bradley2000}. Several studies using a picture-viewing paradigm have shown that EDA is highly correlated with self-reported emotional arousal \citep{Lang1993}. That is, arousing pictures of either positive or negative valence result in increased EDA as compared to low-arousal pictures. This index is affected by the location of recording, as different skin sites are innervated by different distributions of nerve bundles, not all of which are involved in emotional responses. In simple terms, emotional response affects eccrine sweat glands, which are most densely distributed on the palms and soles, nearly four times more so than on the forehead, for example. Sixteen recording sites were explored and compared in a review by \citet{vanDooren2012}, which profiled site-wise responsiveness to emotional inducement (by film clips). Their review illustrates that care must be taken in the choice of the signal \textit{feature} to estimate responsiveness. The authors found also that responses did not show full lateral symmetry, so care must be taken in the decision on which side of the body to record. Picard's Multiple Arousal Theory \citet{Picard2015} suggests an explanation: that different brain areas map to different areas of the body, both contralaterally and ipsilaterally.

\begin{figure}[t]
   \centering
   \includegraphics[scale=0.40]{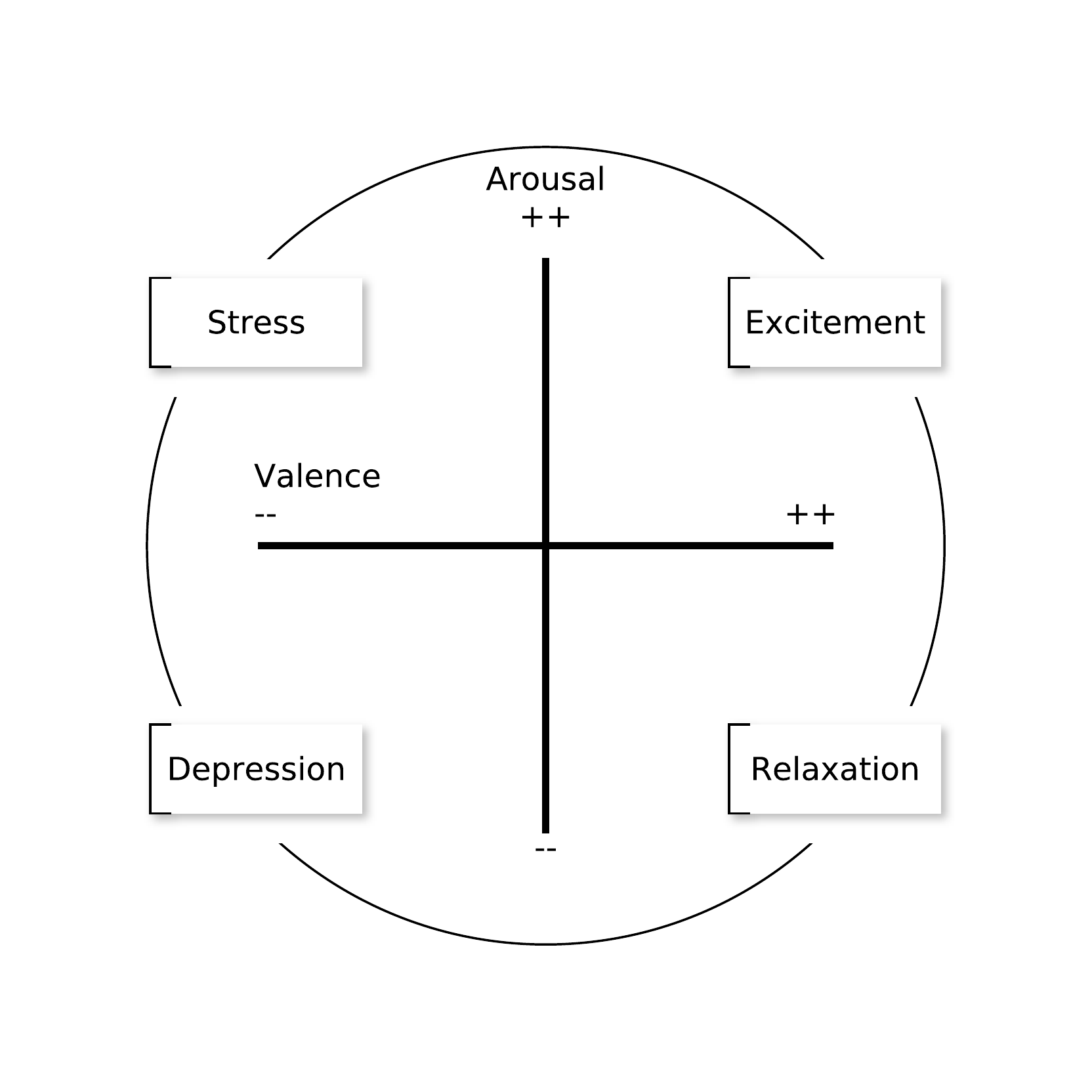}
   \caption{The simple emotional circumplex model, with orthogonal bipolar dimensions of arousal (from alert to lethargic) and valence (pleasant to unpleasant).}
   \label{fig.emocir}
\end{figure}

EDA is a commonly used physiological measure when one is studying HCI experiences (see `Applications', below). The arousal models used in HCI studies are often uni-dimensional and bipolar, and, hence, they can be combined with a dimension of positive--negative valence to give a circumplex model of emotions, as highlighted in Figure~\ref{fig.emocir}.
However, richer models have been proposed, such as the three-system model of arousal \citep{Backs2000}. Indeed, Backs and Boucsein (p. 6) argued that this might be more appropriate for investigating the specific sensitivity of physiological effects in HCI. In brief, this model posits three systems: `affect arousal', `effort', and `preparatory activation', of which \textit{only affect is indexed by EDA}. The areas of the CNS that correspond to these systems are Amygdala, Hippocampus, and Basal Ganglia, respectively. The authors also provided a review demonstrating the sensitivity of EDA in technology interaction studies (p. 16).

%NOT VITAL PICTURE BUT USEFUL. CREATED BY B.C. NO PERMISSION NEEDED

\section{Methods}

% devices
\subsection{EDA Instrumentation}

EDA is a well-established recording method, and numerous devices exist for performing laboratory-grade measurements. These devices usually comprise wired electrodes and often a bulky amplifier, thereby restricting use to controlled environments. Furthermore, electrodes placed on the hand are often very sensitive to motion, thereby requiring the hand to stay quite still.

With the recent increase in the quality and popularity of wearable biosensors, several portable EDA devices have become available. Portability is appealing for both psychological research and clinical use. In psychology, wearable sensors allow experiments to take place in more ecologically valid settings \citep{bete14}, while in health care wearable sensors enable continuous physiological monitoring at a relatively low cost \citep{pante10}.

Non-intrusively measuring EDA in a continuous long-term manner is desirable for many, quite different fields of research and diagnostics. Popular options in this regard are wearable EDA sensors, such as the ring-mounted Moodmetric (Vigofere Ltd., Helsinki, Finland); the wrist-worn E4 (Empatica Inc., Boston, MA, USA); or the edaMove (movisens GmbH, Karlsruhe, Germany), which combines a wrist-worn amplifier with wired electrodes. A recent study addresses the comparability of such a wearable sensor to a laboratory-grade device \citep{Torni2015}.

%recording
\subsection{Recording}

EDA measurement registers the inverse of the electrical resistance `ohm' between two points on the skin -- i.e., the conductivity of the skin in that location, `mho'. The recorded EDA signal has two components. The slowly varying tonic component of the EDA signal represents the current skin conductance level (SCL) and can be influenced by external or internal factors such as dryness of the skin and psychological state. Superimposed on the slow tonic component is a rapidly changing phasic component, skin conductance response (SCR); see, for example, Figure~\ref{fig.scr_overlap}. The spike-like SCR corresponds to sympathetic arousal, resulting from an orienting response to either \textit{specific} environmental stimuli, such as a novel, unexpected, significant, or aversive stimulus, or \textit{non-specific} activation, such as deep breaths and body movements \citep{boucs12, Dawson2000}.

Typically, EDA is recorded non-invasively from the surface of the palms and fingers. Following \citet{Boucsein2012}, we recommend recording from the fingers to the extent that this is possible. Fingers provide good signal characteristics, such as amplitude, and responsiveness of the signal to emotional relevance is well-established. When recording is conducted in situations that demand grasping actions, which could disturb the sensors, the soles of the feet, or the forehead, may be used also \citep{vanDooren2012}.

%preproc and decomposition
\subsection{Preprocessing}

In a typical EDA analysis, the acquired signal is preprocessed and then decomposed into tonic and phasic components -- i.e., SCL and SCRs. The preprocessing is relatively simple: data are down-sampled or low-pass filtered, typically to \textless10~Hz. Electrode displacement tends to generate artefacts, represented by signal discontinuities. These can be detected by a maximum signal-change threshold criterion and handled epoch-wise by rejection or temporal interpolation. For group analysis, the signal should then be standardised or centred. 

\begin{figure}[!htp]
   \centering
   \includegraphics[scale=0.75]{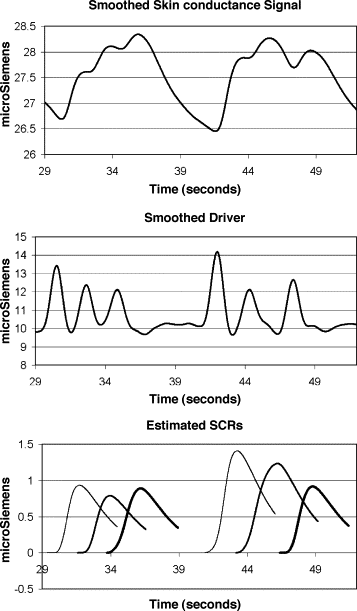}
   \caption{Illustration of SCR overlap, reproduced from \citet{alex05}, with permission. They explain: ``The upper graph shows the smoothed skin conductance signal, with two groups of three overlapping SCRs. The middle graph shows the commuted driver signal, which because of its shorter time-constant has six clearly separate peaks. These separate peaks are used to estimate the individual SCRs shown in the bottom graph.''}
   \label{fig.scr_overlap}
\end{figure}

Signal decomposition can be performed via a number of methods, depending on whether stimulus events also have been recorded. If event times are known, latency-based detection of SCRs can be performed, per \citet{Boucsein2012}. Boucsein and colleagues also define the SCL as the signal in the absence of SCRs; therefore, after SCR detection, SCL can be estimated by subtraction. However, data-driven methods should be preferred, to minimise errors, because SCRs do not follow uniformly from events, and events can occur in rapid succession, causing SCR overlap. 
A classic example is peak-and-trough detection, which is achieved by finding zero-derivative points (where the signal is flat). One can identify SCR features from the trough-to-peak amplitude and latency. This system tends to be inaccurate for stimulus events that overlap -- i.e., that have a shorter inter-stimulus interval than the recovery time of the phasic peak -- because the amplitude of SCRs begins to sum. See Figure~\ref{fig.scr_overlap} for more details.

%nice to have a figure for this, always a confusing concept...permission GRANTED!

\citet{alex05} proposed a method that handles this issue, based on the \textit{deconvolution} of the signal to estimate the driver function from sudomotor nerve activity and the corresponding impulse response function, the latter describing the temporal profile of each impulse of the phasic driver response and used as the deconvolution kernel in the decomposition process.

\begin{figure}[!t]
	\centering
	\includegraphics[scale=0.4]{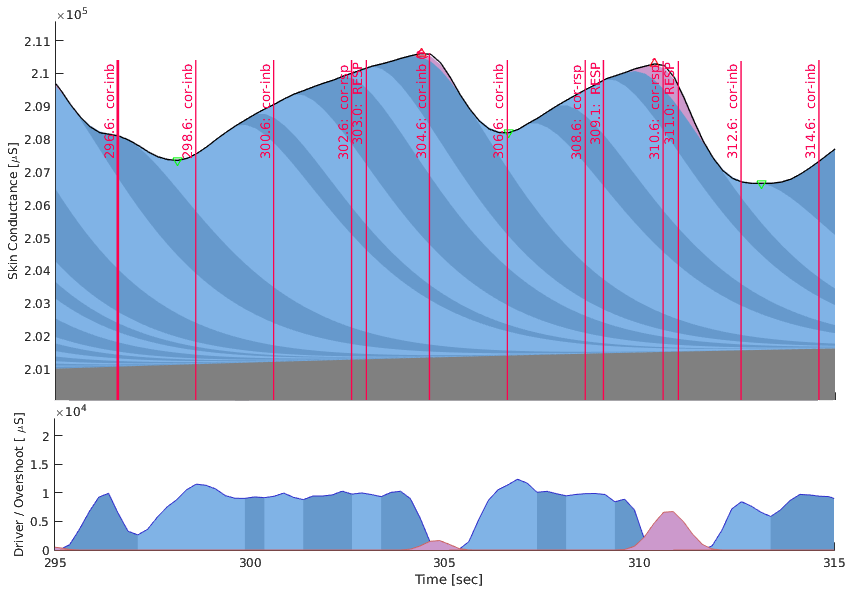}
	\caption{Screenshot from Ledalab. Top panel: 20 seconds of EDA are shown from a recording of a continuous-performance task, with inhibit (labelled `cor-inb') and respond (labelled `cor-rsp') targets shown every $\sim$2 seconds. Response targets (with the subject's responses labelled `RESP') are less frequent in the task so generate a greater EDA response. The grey area indicates SCL, and the coloured area shows SCRs diminishing over time. Bottom panel: Fitting by NND in Ledalab produces an estimate of the SCRs (`Driver', blue) and pore opening components (`Overshoot', pink).}
	\label{fig.leda}
\end{figure}

This method is based on standard deconvolution, which does not account for variations in the SCR shape and can result in a negative driver function when the SCR has a peaked shape. These problems were addressed by \citet{BeneKae2010,Benedek2010}, who introduced two separate solutions: non-negative deconvolution (NND) and continuous decomposition analysis (CDA)\footnote{NND and CDA are implemented as the \textit{Ledalab} toolbox for Matlab.}. Using NND ensures that any negative component of the driver is transformed to a positive `remainder', interpreted as the additional phasic component caused by pore opening. The output of this analysis is depicted in Figure~\ref{fig.leda}. The NND approach was inspired by the poral valve model of EDA, which suggests that peaked SCRs result from additional sweat diffusion caused by pore opening, as illustrated in Figure 1 from \citet{Benedek2010}. They state that
\begin{quote}
	``[i]f sweat ducts are filled to their limits, intraductal pressure will cause a hydraulic-driven diffusion of sweat to the corneum, resulting in a flat SCR. If intraductal pressure exceeds tissue pressure, the distal part of the duct and the pore will eventually open, which results in a peaked SCR''.
\end{quote}

CDA takes a different approach, which ``abandons the concept of single, discrete responses in favour of a continuous measure of phasic activity'' \citep{BeneKae2010}. The latter is, of course, more plausible in a messy biological system. The CDA estimate of the phasic driver can take on negative values, in which case the interpretation is simply that negative values signify quality issues, either in extraction algorithm parameters or in the original data.

For \citet{BeneKae2010}, estimation is a multi-step optimisation process using gradient descent to minimise a compound error consisting of a weighted sum of the negativity and indistinctness of the phasic driver. Indistinctness describes the sharpness of impulses, and negativity represents the number of negative values in the phasic driver.

%a nice picture although not vital. PERMISSION COSTS €900!
%\begin{figure}[ht]
%   \centering
%   \includegraphics[scale=0.5]{figs/Edelberg_poral_valve_model}
%   \caption{Illustration of the poral valve model \citep{Benedek2010}. ``If sweat ducts are filled to their limits, intraductal pressure will cause a hydraulic-driven diffusion of sweat to the corneum, resulting in a flat SCR (A). If intraductal pressure exceeds tissue pressure, the distal part of the duct and the pore will eventually open, which results in a peaked SCR (B)."}
%   \label{fig.poral_valves}
%\end{figure}

%phasic event-related analysis
\subsection{Analysis}

For group phasic analyses, the impulse response function generally should be estimated separately for each participant. The phasic component is then analysed around selected events (if the phasic component was derived by data-driven methods as recommended, without reference to the events, there is the added benefit that a relationship discovered between phasic features and events cannot be an artefact of the feature extraction method). One can do this either by averaging the phasic driver or by calculating a set of phasic features and then performing the analysis in feature space (as in, for example, \citet{khal02}). Commonly used phasic features include the number of significant phasic peaks, the sum of amplitudes of those peaks, the time integral of the phasic response, and the maximum value of phasic activity.

\section{Applications}
EDA has seen application in a host of areas, from research to clinical practice and consumer devices. The number of form factors used in such devices remains relatively limited (they are usually situated on the wrist and fingers), but, as \citet{vanDooren2012} have shown, there are many options for recording sites. Therefore, in line with the application, the reader could conceive of implementing a device in a hat or eyeglasses (to measure forehead EDA), in socks or shoes (to measure foot sole EDA), or in a wrist-worn strap or other clothing items.

There is an extensive body of literature on EDA applications; here, we cite only a few examples.
%Show how the most important indices (from Table 1) are measured with the sensor(s) featured in your chapter, ideally (but optionally) with an illustrative diagram or figure...
% Arousal
% Excitement
% Positive/Negative emotion
% WILL ADD JARI'S EHPG paper about EDA in nuclear power plant operators?

In the area of HCI, EDA is a popular input in helping to classify arousal (usually referring to `affect arousal' \citep{Backs2000}). For example, \citet{Fantato2013} reported on a na{\"i}ve Bayes classifier, which was trained to recognise states of affect arousal from a number of EDA features, on the basis of validated labelling of arousal levels during work-like tasks. Cross-validation testing of these tasks achieved an accuracy level above 90\%. The system was tested also by recording of subjects in a computer-game-like learning environment, where the classifier achieved an accuracy of 69\% for predicting the self-reported emotional arousal of the game. The sensor was the Varioport-ARM device (Becker Meditec, Karlsruhe, Germany).

Studies have shown more specific effects also. \citet{Heiden2005}, studying work done with a computer mouse, found highly significant differences in EDA between conditions that differed in the level of task difficulty.
\citet{Setz2010} compared several classifiers in discriminating between work-like tasks with a baseline cognitive load only and tasks with added stress (considered to be a form of negative affect arousal). The input consisted of 16 EDA features, and the researchers' best-performing classifier (Linear Discriminant Analysis, LDA) achieved an accuracy of 83\%. Their device was an early form of wearable arm-mounted sensor, lab-built and described in the paper referenced above.

%Biofeedback for performance enhancement
\vspace{10 mm}

Offering a final example, we focus on an application that is not usually connected with the workplace. Biofeedback is an increasingly popular application for performance enhancement, and it can be found in such varied contexts as clinical, occupational, and sports scenarios. In clinical biofeedback, the user is trained to respond to a given feature of the real-time signal from a physiological sensor; in this manner, the user can learn to recognise and control the subjective state that corresponds to the feature. With EDA, the feature that needs to be classified might be, for example, the number of significant phasic peaks. In an application, users could learn to recognise the subjective feeling of having more or fewer phasic peaks, then attempt to control their physiological state accordingly.

%recreation
One recreational use of biofeedback involves an affect-based music player, in which concurrently measured biosignals are used to classify the listener's emotional response as the music is playing. The efficacy of such a system for inducing target moods has been demonstrated in an ecologically valid office setting, although with only a small sample size, \emph{N}=10 \citep{VanderZwaag2013}.

%clinical
\citet{OConnell2008} demonstrated the `Self-Alert Training' (SAT) system for EDA biofeedback, to modulate attention via arousal level. This software was validated with a group of 23 neurologically healthy participants, each of whom received brief (30--40-minute) biofeedback training sandwiched between two sustained attention to response task (SART) tests. Half of the participants were given placebo training. Analysis indicated that the SAT group
\begin{itemize}
	\item significantly reduced their number of commission errors (a measure of response inhibition), while the placebo control group did not;
    \item maintained consistent response time variability (RTV -- an inverse measure of sustained attention) after training, whereas the placebo group shows a significant increase in RTV; and
    \item increased in arousal (SCR amplitude in response to cues) after training, while the placebo group's corresponding figures significantly decreased.
\end{itemize}
The last of these findings indicates that the short training period was enough to enable participants to counter whatever effects of fatigue and cue exposure had caused the reduction of arousal in the placebo group. This is important for the domain of safety-critical operator work in an HCI setting, where the effect of brief periods of activity to boost vigilance and alertness can be considered a valuable option for reducing human error. Such systems can now be implemented at low cost, as sensor devices are becoming robust, lightweight, and wearable, and interfaces are available for mobile platforms such as smartphones.

\section{Conclusion}

Electrodermal activity is a reliable, interpretable, and simple-to-use measure that has seen many applications in various domains. Therefore, it is an excellent choice for an introduction to the psychophysiological method and a highly suitable tool for making inferences about sympathetic nervous system activity. In addition, EDA aids in providing valuable context for other physiological signals in multimodal applications.

\bibliographystyle{plainnat}
\bibliography{ch2_eda_bib}

\end{document}